\documentclass[aps,prl,twocolumn,groupedaddress]{revtex4}

\usepackage{epstopdf}
\usepackage{graphicx}
\usepackage{rotating}

\def\beq{\begin{equation}}
\def\eeq{\end{equation}}
\def\bea{\begin{eqnarray}}
\def\eea{\end{eqnarray}}  

\def\eq#1{{Eq.~(\ref{#1})}}
\def\fig#1{{Fig.~\ref{#1}}}

\newcommand{\SP}{\langle \mid S \mid^2 \rangle}

\newcommand{\Lb}{\left(}
\newcommand{\Rb}{\right)}
\parskip=\itemsep

\newcommand{\A}{{\cal A}}

\begin{document}


\title{A Study of Soft Interactions at Ultra High Energies}
\thanks{Talk given by U. Maor}

\author{E. Gotsman}
\email[]{Email: gotsman@post.tau.ac.il}
\author{E. Levin}
\email[]{Email: leving@post.tau.ac.il}
\author{U. Maor}
\email[]{Email: maor@post.tau.ac.il}
\affiliation{Department of Particle Physics, School of Physics and Astronomy\\
Raymond and Beverly Sackler Faculty of Exact Science\\
Tel Aviv University, Tel Aviv, 69978, Israel}


\begin{abstract} 
We present and discuss our recent study of an eikonal two channel model, 
in which we reproduce  
the soft total, integrated elastic and diffractive cross sections,
and the corresponding forward differential slopes in the ISR-Tevatron
energy range.
Our study is extended to provide predictions at the LHC and Cosmic Rays energies. 
These are utilized to assess the role of unitarity at ultra high energies, 
as well as predict the implied survival probability of exclusive diffractive 
central production of a light Higgs. 
Our approach is critically examined so as to estimate the margins of error 
of the calculated survival probability for diffractive Higgs production.
 
\end{abstract}
\maketitle
\section{Introduction}
The search for unambiguous s-channel unitarity signatures 
in ultra high energies soft hadronic scattering, is two folded: 
\newline
On the one hand,  
this is a fundamental issue on which we have only limited 
information from the ISR-Tevatron experiments. 
The only direct indication we have on the importance of 
unitarity considerations, derives from the observation that soft 
diffraction cross sections, essentially SD (single diffraction), have a
much milder energy 
dependence than the seemingly similar, elastic cross sections. 
Enforcing unitarity constraints is a model dependent procedure. 
Thus,  reliable modeling is essential  for the execution of our 
study, leading to predictions of interest for LHC and AUGER experiments.
\newline
On the other hand, 
unitarity considerations in soft scattering are instrumental for the
assessment of inelastic hard diffraction rates, specifically,  
diffractive Higgs production at the LHC. Preliminary information on 
the importance and method of this calculation has been acquired in the 
study of hard diffractive di-jets at the Tevatron\cite{heralhc}, leading 
to first generation estimates of the corresponding survival probabilities.
\par
This presentation is based on our recent paper\cite{GLM07}, 
which utilizes
the GLM model\cite{heralhc,1CH,2CH,SP2CH,GKLMP,GKLM} where we numerically 
solve the $s$-channel unitarity equation in an eikonal model.
Our updated results, in the ISR-Tevatron range, were 
obtained from an improved two channel model calculations.
The specific objectives of our study, based on the above, were:
\newline
1) To reproduce the total, integrated elastic and diffractive 
cross sections and corresponding forward differential slopes
in the ISR-Tevatron energy range, 
and to obtain predictions for these observables 
at LHC and Cosmic Rays energies. 
\newline
2) To calculate the survival probabilities of inelastic hard 
diffractive processes\cite{Bj,GLM1}. 
This requires precise knowledge of the soft elastic and 
diffractive scattering amplitudes of the initial hadronic projectiles. 
As we noted, it is of particular importance
for the assessment of the discovery potential for LHC Higgs production in an
exclusive central diffractive process.  
\newline
3) Some of the fundamental consequences of
s-channel unitarity in the high energy limit are not clear, as yet.
We examin the approach of the scattering amplitudes 
to the black disc bound.
\newline
4) We  estimate the margin of error of our predicted 
survival probabilities,     based on 
a critical analysis of our model. 
\section{The GLM Model}
The main assumption of the two channel GLM model is 
that hadrons are the 
correct degrees of freedom at high energies, diagonalizing the scattering matrix.
In this Good-Walker type formalism, 
diffractively produced hadrons at a given vertex are
considered as a single hadronic state
described by the wave function $\Psi_D$, which is orthonormal
to the wave function $\Psi_h$ of the incoming hadron, $<\Psi_h|\Psi_D>=0 $.
We introduce two wave functions $\Psi_1$ and $\Psi_2$ which diagonalize the
2x2 interaction matrix ${\bf T}$
\beq \label{2CHM}
A_{i,k}^{i',k'}=<\Psi_{i}\,\Psi_{k}|\mathbf{T}|\Psi_{i'}\,\Psi_{k'}>=
A_{i,k}\,\delta_{i,i'}\,\delta_{k,k'}.
\eeq
In this representation the observed states are written 
\beq \label{2CHM31}
\Psi_h=\alpha\,\Psi_1+\beta\,\Psi_2\,,
\eeq
\beq \label {2CHM32}
\Psi_D=-\beta\,\Psi_1+\alpha \,\Psi_2\,,
\eeq
where, $\alpha^2+\beta^2=1$.
\par
Using \eq{2CHM} we can rewrite the unitarity equations 
\beq \label{UNIT}
Im\,A_{i,k}\left(s,b\right)=|A_{i,k}\left(s,b\right)|^2
+G^{in}_{i,k}(s,b),
\eeq
where $G^{in}_{i,k}$ is the summed probability for
all non diffractive inelastic processes induced by the initial $(i,k)$ states. 
The simple solution to \eq{UNIT} has the form obtained in a single channel 
formalism\cite{1CH}, 
\beq \label{2CHM1}
A_{i,k}(s,b)=i \Lb 1 -\exp\Lb - \frac{\Omega_{i,k}(s,b)}{2}\Rb\Rb, 
\eeq
\beq \label{2CHM2}
G^{in}_{i,k}(s,b)=1-\exp\Lb - \Omega_{i,k}(s,b)\Rb.
\eeq
From \eq{2CHM2} we deduce the probability that the initial projectiles
$(i,k)$ reach the final state interaction unchanged, regardless of the initial
state re-scatterings, is given by
$P^S_{i,k}=\exp \Lb - \Omega_{i,k}(s,b) \Rb$.
\par
In general, we have to consider four possible $(i,k)$ 
re-scattering options.
For initial $p$-$p$ (or $\bar p$-$p$) the two quasi-elastic amplitudes are
equal $A_{1,2}=A_{2,1}$, and we have three re-scattering amplitudes. The 
corresponding elastic, SD and DD amplitudes are  
\beq \label{EL}
a_{el}(s,b)=
i\{\alpha^4A_{1,1}+2\alpha^2\beta^2A_{1,2}+\beta^4\A_{2,2}\},
\eeq
\beq \label{SD}
a_{sd}(s,b)=
i\alpha\beta\{-\alpha^2A_{1,1}+(\alpha^2-\beta^2)A_{1,2}+\beta^2A_{2,2}\},
\eeq
\beq \label{DD}
a_{dd}=
i\alpha^2\beta^2\{A_{1,1}-2A_{1,2}+A_{2,2}\}.
\eeq
Adjusted parameters are introduced  to obtain explicit
expressions for the opacities $\Omega_{i,k}(s,b)$. 
\par
In the following we shall consider Regge and non Regge options for the
dynamics of interest. We use a simple general form  
for the input opacities,
\beq \label{omega}
\Omega_{i,k}\Lb s,b \Rb = \nu_{i,k}\Lb s \Rb \Gamma\Lb s,b \Rb. 
\eeq
\beq
\nu_{i,k}\Lb s \Rb = \sigma^0_{i,k}\,\Lb \frac{s}{s_0}\Rb^{\Delta}.
\eeq
The input b-profiles $\Gamma_{i,k} \Lb s,b \Rb$ are assumed to be Gaussians
in b, corresponding to exponential differential cross sections in t-space, 
\beq
\Gamma_{i,k} \Lb s,b \Rb = \,\,\frac{1}{\pi R^2_{i,k}\Lb s \Rb}
\exp \Lb - \frac{b^2}{ R^2_{i,k}\Lb s \Rb}\Rb,
\eeq
\beq \label{radius}
R^2_{i,k}\Lb s \Rb = R^2_{0;i,k}+4C ln(s/s_0).
\eeq
$R^2_{0;1,2}=\frac{1}{2}R^2_{0;1,1}$ and $R^2_{0;2,2}=0$. 
Our parametrization is compatible with, but not exclusive to, 
a Regge type input.
\section{Fits and Predictions}
We have studied three models, with different parameterizations 
of $\Omega_{i,k}$, which were adjusted to the ISR-Tevatron   
experimental data base, specified above.
Note that the fit has, in addition to the
contribution in the form of \eq{omega}, also a secondary Regge sector
(see Ref.\cite{1CH,2CH}).
This is necessary, as the data base contains a relatively  small number
of experimental high energy measured values, which are independent of 
the Regge contribution. We do not quote the values of the Regge parameters, 
as the goal of this paper is to obtain predictions in the
LHC and Cosmic Rays energy range.
At W=1800$GeV$  the Regge sector 
contribution is less than 1$\%$. However, it is essential at the ISR energies.  
\par
Model A is a simplified two amplitude version of the two channel model, 
in which we assume that $\sigma_{dd}$ is small enough to be neglected. 
As such, this model breaks Regge factorization. 
The model was presented and discussed in Ref.\cite{2CH}.
The parameters of Model A were obtained from a fit to a 55 
experimental data points base and are listed in Table 1 with 
a corresponding $\chi^2/(d.o.f)$ of 1.50. Note that in Model A 
the (1,1) amplitude corresponds to $\Omega_{1,1}$, while the (1,2) 
amplitude corresponds to 
$\Delta \Omega=\Omega_{1,1}-\Omega_{1,2}$. See Ref.\cite{2CH}. 
\begin{footnotesize}
\begin{table}
\begin{tabular}{|l|l|l|c|}
\hline
&Model A &  Model B(1) &  Model B(2) \\ \hline
$\Delta$ & 0.126 & 0.150 & 0.150 \\
$\beta$ & 0.464 & 0.526 & 0.776 \\
$R^{2}_{0;1,1}$ & 16.34 $GeV^{-2}$ & 20.80 $GeV^{-2}$ &20.83 $GeV^{-2}$\\
$C$&0.200 $GeV^{-2}$&0.184 $GeV^{-2}$&0.173 $GeV^{-2}$\\
$\sigma^0_{1,1}$&12.99 $GeV^{-2}$&4.84 $GeV^{-2}$&9.22 $GeV^{-2}$\\
$\sigma^0_{2,2}$& N/A &4006.9 $GeV^{-2}$&3503.5 $GeV^{-2}$\\
$\sigma^0_{1,2}$& 145.6$GeV^{-2}$ &139.3 $GeV^{-2}$&6.5 $GeV^{-2}$\\ \hline
\end{tabular}
\caption{Fitted parameters for Models A, B(1) and B(2).}
\end{table}
\end{footnotesize}
\par
Model B denotes our three amplitude model where the 5 published 
DD cross section points\cite{DDD} are contained in the fitted data base.
The three opacities are taken to be Gaussians in $b$. 
If we assume the soft Pomeron to
be a simple J pole, its coupling factorization implies 
$\sigma^0_{1,2} = \sqrt{\sigma^0_{1,1}\times \sigma^0_{2,2}}$. 
We denote this Model B(1). The fit obtained is not satisfactory, 
with a $\chi^2/(d.o.f.)$=2.30.
\par
We have, also, studied Model B(2) in which coupling factorization 
is not assumed. Accordingly, $\sigma^0_{1,1}$, 
$\sigma^0_{1,2}$ and $\sigma^0_{2,2}$ 
are independent fitted parameters of the model. 
The model with a $\chi^2/(d.o.f.)$ = 1.25, provides a
very good reproduction of our data base.
In Model B(2) the leading t channel exchange is not a simple J pole. 
It is compatible with a model\cite{SAT} we have suggested
a while ago in which the soft Pomeron dominated photo and low $Q^2$ DIS, is
perceived as the saturated soft (low $Q^2$) limit of the hard Pomeron dominated 
(high $Q^2$) hard DIS. A major deficiency of Model B(2) is that it predicts 
dips in $\frac{d\sigma_{el}}{dt}$ at small $t$ values, which are not
observed 
experimentally. This problem is common to all eikonal models  which  assume 
Gaussian b-profiles. Consequently, Model B(2) is valid only in the 
narrow forward $t$ cone, where it reproduces  approximately
85$\%$ of the 
overall data very well. 
We shall discuss this problem in some detail in
the Discussion 
Section. 
\begin{footnotesize}
\begin{table}
\begin{tabular}{|l|l|l|l|l|l|c|c|c|c|}
\hline
&         &                       &
&         &                       &
&         &                       \\
$\sqrt{s}$ & $\sigma_{tot} $ & $  \sigma_{el} $  & $ \sigma_{sd} $&
$\sigma_{dd}$& $B_{el}$ & $R_{el}$ & $R_D$
& $\frac {\sigma_{diff}} {\sigma_{el}}$ \\
TeV & mb & mb & mb & mb & $GeV^{-2}$&  &  & \\ \hline
1.8 & 78.0 & 16.3 & 9.6 & 3.8 & 16.8 & 0.21 & 0.38 &0.83\\
14 & 110.5 & 25.3 & 11.6 & 4.9 & 20.5 & 0.23 & 0.38 &0.65 \\
30 & 124.8 & 29.7 & 12.2 & 5.3 & 22.0 & 0.24 & 0.38 &  0.59\\
60 & 139.0 & 34.3 & 12.7 & 5.7 & 23.4 & 0.25 & 0.38 &0.54 \\
120 & 154.0 & 39.6 & 13.2 & 6.1 & 24.9 & 0.26 & 0.38 & 0.49\\
250 & 172.0 & 45.9 & 13.6 & 6.6 & 26.5 & 0.27 & 0.38 & 0.44 \\
500 & 190.0 & 52.7 & 14.0 & 7.0 & 28.1 & 0.28 & 0.39 & 0.40 \\
1000 & 209.0 & 60.2 & 14.3 & 7.4 & 29.8 & 0.29 & 0.39 & 0.10 \\
$10^{11}$ & 1070.0 & 451.2 & 21.6 & 19.5& 109.9 & 0.42 & 0.46 & 0.09\\
1.22\,$10^{19}$ & 1970.0 & 871.4 & 25.5 &27.7 & 202.6 & 0.44 & 0.47 & 0.06\\
(Planck) & & & & & & & & \\
\hline
\end{tabular}
\caption{Cross sections and elastic slope in Model B(2).}
\end{table}
\end{footnotesize}
\par
Model B(2) cross section and slope 
predictions at ultra high energies are 
summarized in Table 2. Note that $R_{el}=\sigma_{el}/\sigma_{tot}$ 
and $R_{D}=(\sigma_{el}+\sigma_{diff})/\sigma_{tot}$.
At LHC (W=14 $TeV$) our predicted cross sections are:
$\sigma_{tot}=110.5\,mb$, $\sigma_{el}=25.3\,mb$, $\sigma_{sd}=11.6 \,mb$
and $\sigma_{dd}=4.9\,mb$. 
These predictions are slightly
higher than those obtained\cite{2CH} in Model A.
The corresponding forward slopes are: $B_{el}=20.5 \,GeV^{-2}$,
$B_{sd}=15.9\,GeV^{-2}$ and $B_{dd}=13.5\,GeV^{-2}$. We calculate, 
also, $\rho=0.125$.
The calculations of $B_{sd}$, $B_{dd}$ and $\rho$
were executed with the fitted parameters of the model.
For the record we have checked that we reproduce also the UA4, CDF and E710 
$B_{sd}$ and $\rho$ data points. 
\section{Survival probabilities}
In the following we shall limit our discussion to the survival probability
of Higgs production in an exclusive
central diffractive process, calculated in our model.
For a general review see 
Ref.\cite{heralhc}. 
\par
In our model we assume an input Gaussian $b$-dependence
also for the hard diffractive amplitude of interest. Its 
input, when convoluted with the soft (i,k) channel, is 
\begin{equation}\label{3.6}
{\Omega_{i,k}^H}={\nu_{i,k}^H(s)} \Gamma_{i,k}^H(b),
\end{equation}
\begin{equation}\label{3.7}
\nu_{i,k}^H=\sigma_{i,k}^{H0}(\frac{s}{s})^{\Delta_H},
\end{equation}
\begin{equation}\label{3.8}
\Gamma_{i,k}^H(b)=\frac{1}{\pi {R_{i,k}^H}^2}\,e^{-\frac{\,b^2}{{R_{i,k}^H}^2}}.
\end{equation}
\begin{figure}
\includegraphics[width=65mm]{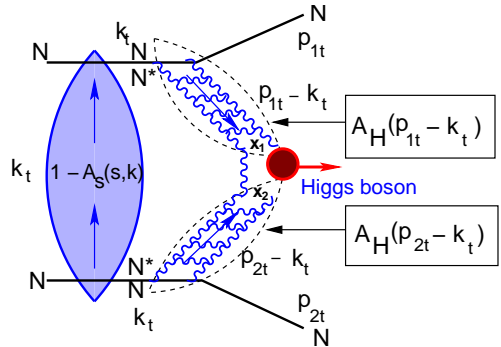}
\caption{\label{sp-dia}Survival probability for exclusive central diffractive
production of the Higgs boson}
\end{figure}
\par
The structure of the survival probability
expression is shown in \fig{sp-dia}.
The corresponding general formulae for the calculation 
of the survival probability for
diffractive Higgs boson production have been discussed
in Refs.\cite{SP2CH,heralhc,GKLMP}. Accordingly,
\beq \label{SP}
\SP=\frac{N(s)}{D(s)},
\eeq
\bea
&N(s) = \int d^2\,b_1\,d^2\,b_2
\{A_H(s,b_1)\,A_H(s,b_2)
\nonumber \\&
(1-A_S (s,(\mathbf{b}_1+\mathbf{b}_2 )))\}^2,
\label{SP1}\\
&D(s) = \int\,d^2\,b_1\,d^2\,b_2
\{A_H(s,b_1)\,A_H(s,b_2)\}^2.
\label{SP2}
\eea
$A_s$ denotes the soft strong interaction amplitude given
by \eq{2CHM1}.
Using \eq{EL}-\eq{DD}, the integrands of
\eq{SP1} and \eq{SP2} are reduced by eliminating common $s$-dependent
expressions.
\bea
&&N(s) =
\int\,d^2 b_1 d^2 b_2
\{(1 - a_{el}(s,b))A^{pp}_H(b_1) A^{pp}_H(b_2)
\nonumber \\&&
 - a_{sd}(s,b)\Lb A^{pd}_H(b_1) A^{pp}_H(b_2) +
A^{pp}_H(b_1) A^{pd}_H(b_2) \Rb
\nonumber \\&& - a_{dd}(s,b) A^{pd}_H(b_1) A^{pd}_H(b_2)\}^2
\label{SP3},
\eea
\beq \label{SP4}
D = \int d^2 b_1 d^2 b_2
\{A^{pp}_H(b_1) A^{pp}_H(b_2)\}^2.
\eeq
\par
Following Refs.\cite{heralhc,GLM07} we introduce
two hard $b$-profiles
\bea
A^{pp}_H(b) &=&
\frac{V_{p \to p}}{2 \pi B_{el}^H} \exp \Lb-\frac{b^2}{2\,B_{el}^H} \Rb,
\label{2C10}\\
A^{pd}_H(b) &=& \frac{V_{p \to d}}{2 \pi
B_{in}^H} \exp \Lb -\frac{b^2}{2 B_{in}^H}\Rb.
\label{2C11}
\eea
The hard radii ${R_{i,k}^H}^2$ and cross section coefficients $V_{p \to p}$ 
and $V_{p \to d}$ 
are constants derived from HERA $J/\Psi$ elastic and inelastic
photo and DIS production\cite{KOTE,PSISL} (see, also, Ref.\cite{GKLMP}).
$B_{el}^H=3.6 GeV^{-2}$, $B_{in}^H=1 GeV^{-2}$, 
$V_{p \to p}=\sqrt{3}$ and $V_{p \to d}=1$. 
have been taken from the experimental HERA data on
$J/\Psi$ production in HERA\cite{KOTE,PSISL}.
\par
Using \eq{SP}-\eq{SP4} we calculate the survival probability $S^2_H$ for
exclusive Higgs production in central diffraction. 
$S^2_H$ has been
calculated\cite{heralhc} in the two amplitude Model A. The resulting
$S^2_H\,=\,0.027$ is essentially the same as
the predictions of KMR\cite{KKMR}.
Our present results,
obtained in the three amplitude B Models, indicate 
a reduction of the output value of
$S^2_H$. Its LHC value in Model B(1) is 0.02, and in
Model B(2) it is 0.007. We note that,
our Model B(1) result is compatible with the result of Ref.\cite{KKMR}. 
We shall return to this issue in the Discussion Section.
\section{Amplitude Analysis}
The   basic amplitudes of the GLM two channel model 
are $A_{1,1}$, $A_{1,2}$ and $A_{2,2}$,
whose $b$ structure is specified in \eq{2CHM1}). 
These are the building blocks
with which we construct $a_{el}$, $a_{sd}$ 
and $a_{dd}$ (\eq{EL}-\eq{DD}). 
The $A_{i,k}$ amplitudes are bounded by the black
disc unitarity bound of unity.
Checking Table 1, it is evident that in both Model B(1) and B(2) 
$\Omega_{2,2}$ is much larger than the other two fitted opacities. 
As a consequence, the amplitude 
$A_{2,2}(s,b)$ reaches the unitarity bound of 
unity at  low energies. Similarly, the output amplitude   
$A_{1,2}(s,b)$ of Model A reaches unity at approximately LHC energy. 
The observation that one, or even two, of our $A_{i,k}(s,b)$=1 does not 
imply that the elastic scattering amplitude 
has reached the unitarity bound at these $(s,b)$ values.
$a_{el}(s,b)$ reaches the black disc bound when, and only when, 
$A_{1,1}(s,b)$=$A_{1,2}(s,b)$=$A_{2,2}(s,b)$=1. In such a case 
we also obtain, that $a_{sd}(s,b)$=$a_{dd}(s,b)$=0.
This result is independent of the fitted value of $\beta$.
\par
\begin{figure}
\includegraphics[width=80mm]{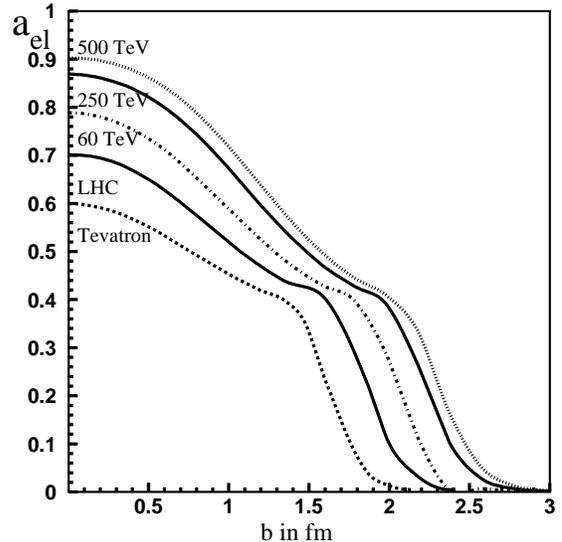}
\caption{b dependence of $a_{el}$ in Model B(2) at different energies}
\label{ampel}
\end{figure}

Model B(2) predictions of $a_{el}$ over a wide range of energies 
are presented in \fig{ampel}.
A fundamental feature of Models A, B(1) and B(2) is that
$a_{el}$ approaches the black disc bound at $b=0$ very slowly,
reaching the bound at energies higher than the GZK knee cutoff. 
If correct, this feature implies that $a_{el}$ does not reach the 
black disc bound over the entire accessible spectrum of Cosmic Rays energies, 
even though it gets monotonically darker. 
\par
The explanation of 
this behavior, in our presentation, is simple.
Checking the values of $\beta$ and $\sigma_{i,k}^0$
corresponding to the 3 models (see Table 1), we note that 
$\Omega_{1,1}$ is smaller by 1-3 orders of magnitude relative to 
$\Omega_{2,2}$ ($\Omega_{1,2}$ in Model A). 
The consequent $a_{el}$ can reach the black disc bound only when $\Omega_{1,1}$
is large enough so that $A_{1,1}$ approaches unity. $\Omega_{1,1}$ grows 
slowly like $W^{0.3}$ (modulu $ln W$). Hence, 
the slow approach of $a_{el}$ toward the black disc bound.  
This result is incompatible with the output of Ref.\cite{KKMR} in which 
$a_{el}$ reaches the black disc bound approximately at the LHC. In our 
presentation it implies that unlike our models, in the KMR model there 
is relatively small variance in the weights of the 3 components 
of the proton wave function. 
\par
A consequence of the input $\Omega_{i,k}$ being large at small $b$, 
is that $P_{i,k}^S(s,b)$ is very small at $b$ = 0 and 
monotonically approaches its limiting value of 1, in the high $b$ limit.
As a result, given a diffractive (non screened) input, 
its output (screened)
amplitude is peripheral in $b$. This is a general feature, common to 
all eikonal models regardless of their b-profiles details. The same is, 
true, also, with regard to diffractive Good-Walker 
channels, which are contained in $\Omega_{i,k}$. 
This implies a non trivial $t$ dependence of
$d\sigma_{diff}(M^2_{diff})/dt$ in the diffractive channels.
These qualitative features are induced by Model A, B(1) and B(2), 
even though their detailed behavior are not identical. Given the 
deficiencies of our b-profiles, we refrain from giving any specific 
predictions beside the general observation stated above. 
\par
The general behavior indicated above 
becomes more extreme at ultra high energies, 
when $a_{el}$ continues to expand and gets darker. 
Consequently, the inelastic diffractive channels becomes more and more 
peripheral and relatively smaller when compared with the elastic channel. 
At the extreme, when $a_{el}(s,b)$ = 1, $a_{sd}\,=\,a_{dd}\,=\,0$. 
We demonstrate this feature and its consequence at 
the Planck mass in \fig{ampel}.
As the black core of $a_{el}$ expands, the
difference between Models A, B(1) and B(2), considered in this paper,
diminishes, being confined to the narrow $b$ tail where $a_{el}(s,b)<1$.
The above observations may be of interest in the analysis of Cosmic Ray
experiments.
\section{Discussion}
\begin{figure}
\includegraphics[width=70mm]{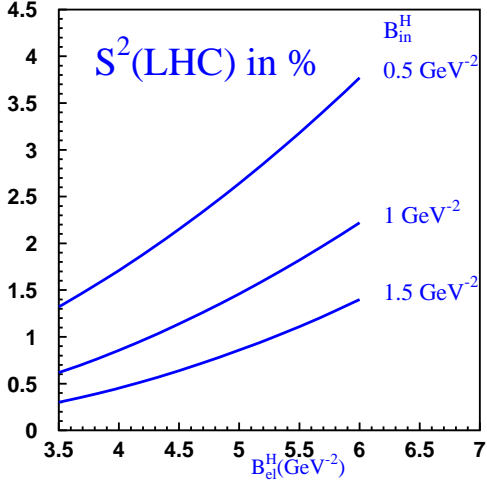}
\caption{ The dependence of $S^2$ at the LHC on $B^H_{el}$ and $ B^H_{in}$,
the slopes for the hard cross sections.}
\label{BH}
\end{figure}

It is interesting to compare our model and its output with a different  
eikonal model recently proposed by KMR\cite{KMR} extending earlier 
versions\cite{KKMR}.
The two models were constructed with very similar objectives but are 
fundamentally different in their conceptual theoretical input, data analysis 
and output results.
\newline
1) The input of KMR is a conventional Regge model in which high 
mass diffraction, initiated by Pomeron enhanced diagrams, is included.
GLM is a phenomenological parametrization in which we assume 
diffraction to be strictly Good-Walker type, with no high mass 
diffraction distinction.
We formulate our 
input in a general form consistent with Regge, but not exclusively so. 
Our statistically preferred non factorizeable Model B(2) is compatible with a 
partonic interpretation 
which considers the soft "Pomeron" to be a low $Q^2$  
high density limit of the hard Pomeron\cite{SAT}. 
The GLM "Pomeron" is not a Regge simple J-pole, it does not include 
Pomeron enhanced diagrams, which are essential in the construction of KMR. 
\newline
2) Since multi-Pomeron vertices are included in KMR, they had to fix 
$\alpha^{\prime}=0$. In order to maintain the experimentally observed 
forward $t$-cone shrinkage, they constructed a high absorption eikonal model 
in which the input is non conventional $\Delta=0.55$. With this input, KMR 
obtain an approximate DL behavior\cite{DL} in the ISR-Tevatron range. 
However, at higher energies their effective $\Delta$ becomes monotonically 
smaller (its value in the Tevatron-LHC range is reduced to 0.04) which results  
in a very slow rise of $\sigma_{tot}$ and $\sigma_{el}$. GLM is a weak screening 
eikonal model. 
Its fitted input is $\Delta=0.15$ and $C=\alpha^{\prime}=0.17$. With this 
input, GLM total cross sections are compatible with DL over the wide 
ISR-GZK range.
\newline 
3) The goal of both GLM and KMR is to adjust the model parameters of their 
vacuum t exchange "Pomeron" input, so as to predict and calculate observables and 
factors of interest at the LHC and Cosmic Rays. 
Both models adjust more than 10 free 
parameters.  Only  CERN-UA4 and Tevatron energies are sufficiently high to 
justify neglecting the contribution of the secondary Regge sector. This 
limited data base is not sufficient to adjust the "Pomeron" free 
parameters. GLM chose, therefore, to construct a model containing also the 
secondary Regge sector and fit the extended data base spanning the ISR-Tevatron 
energy range. KMR constrain their parameter adjustment to the small data base of 
the highest energies. In our opinion the KMR procedure is not adequate. Indeed,  
their reconstruction of $\frac{d\sigma_{el}}{dt}$ at the 3 highest available 
energies is remarkably similar to a fit they made a few years ago with different 
parameters, notably a conventional $\Delta$ input.
\newline
4) GLM and KMR determine their input opacities in completely different procedures 
which define their (different) data bases. GLM approach is that a model 
which 
takes into account diffractive re-scatterings of the initial 
projectiles has to reconstruct properly the diffractive cross sections, which 
are, thus, included in its fitted data base. KMR goal is to reconstruct 
$a_{el}(s,b)$ for which the diffractive components are needed. To this end they 
fit $\frac{d\sigma_{el}}{dt}$ neglecting an explicit fit of the 
diffractive channels. Obviously, combining both GLM and KMR data bases is 
advisable. Regretfully, we were unable to obtain good simultaneous
reproduction of 
such an extended data base. The question, is thus, which model provides
a better 
approximation for the input opacities. 
\newline
5) The b-distributions of $a_{el}(s,b)$ in GLM 
are significantly different from KMR. GLM obtain a  
relatively wide b distribution compared with a narrower one in KMR.  
$a_{el}(s,b=0)$ in KMR is consistently larger than in GLM,    
approaching the black disc bound much faster than in GLM. Regardless of these  
differences, the corresponding values of $\sigma_{tot}$ and $\sigma_{el}$ in both 
models in the UA4-Tevatron range are compatible. 
Such compatibility can exist only over a  
relatively narrow energy band and it 
cannot persist over a wide energy range. Indeed,  
the two models have different LHC and Cosmic Rays predictions, which 
hopefully will be tested soon.
Our inability to reproduce
$\frac{d\sigma_{el}}{dt}$ outside the narrow forward $t$ 
cone implies a deficiency
in our $a_{el}$ at large $b$. We are not clear if this
deficiency is 
reponsible for the small $S_H^2$ obtained in our Model B(2). 
Note, that even though our 
factorizable Model B(1) has the same feature of spurious dips 
outside the very forward at $t$ cone, 
its predicted $S_H^2$ is 0.02 which is compatible with KMR.
\newline
6) In our opinion, the data adjustment procedure adopted by KMR are not adequate.  
Our approach is to quantify our fit by minimizing its $\chi^2$. KMR 
reject any statistical approach to their data analysis. 
They tune many of their parameters by eye and refrain from a 
quantified assessment of their output. The difference between the procedures 
adopted by the two groups is cardinal, as one is unable to make a 
systemic evaluation of the KMR output. 
\newline
7) The difference between the $S_H^2$ predictions of GLM and KMR are intriguing 
and reflect the sensitivity of $S_H^2$ to each model input.
$S_H^2$ is calculated as a convolution of the hard amplitude for Higgs production
and the soft probability $P^S_{i,k}(s,b)$. The hard amplitude features needed for
this calculation in our model are
the hard slopes $B^H_{el}$, $B^H_{in}$ and cross section coefficients 
$V^2_{p \to p}$ $V^2_{p \to d}$, 
determined from the HERA measured\cite{KOTE,PSISL} in
$J/\Psi$ photo and DIS elastic and inelastic production.
Our sensitivity to these parameters is shown in \fig{BH}.
Note that when we change the value of $B_{in}^H$, we keep the ratio
$V^2_{p \to d}/B^H_{in}$ unchanged. Doing so we do not change the
cross section of the reaction
$\gamma  + p \to J/\Psi + X \mbox{(M $\leq$ 1.6 GeV)}$.
KMR calculation is simpler in as much as they consider just the elastic hard
slope. In our opinion there is a gap between the sophistication of KMR soft 
model and the simplicity of their hard approximation. Since $S_H^2$ is obtained 
from a convolution of the two terms it is not clear what is the contribution of 
KMR hard term to the margin of error in their calculation of $S_H^2$.  
\par
KMR estimate their margin of uncertainty to be a factor 
of 2.5. Since our uncertainty derives from similar, though not identical,  
sources, our assessment is similar. 
As we saw, both GLM and KMR models are partially deficient. We noted that 
these are based on the different conceptual constructions and  
data analysis procedures of the two models. 
A discrimination between the two models depends on experimental 
results which are expected to become 
available within the next few years. In the following we list a few:   
\newline
1) GLM predictions for $\sigma_{tot}$ and $\sigma_{el}$ at the LHC are 20$\%$
higher than the corresponding KMR values. This is a fundamental difference since 
the output energy dependence of GLM, which is a weak screening model, 
is compatible with an effective $\Delta=0.08$ all through the 
Tevatron-GZK energy range. 
In the KMR model the effective $\Delta$ is reduced rapidly due to the 
very strong screening which is inherent to this model. 
Hence, the KMR cross sections grow very moderately above the Tevatron energy. 
\newline
2) The difference between the two models becomes more distinguished at 
Cosmic Rays energies. This may be checked by the Auger experiments 
where we expect soon some cross section results at  
energies spanning up to W = 100-150 $TeV$.
\newline
3) A basic feature particular to the KMR model is 
a contribution to diffraction which originates 
from the Pomeron induced diagrams which are not contained in GLM. 
As a result, both $\sigma_{sd}$ and $\sigma_{dd}$ predicted by KMR  
are larger than GLM. These differences are very significant for the DD channel 
where the KMR prediction at LHC is almost a factor of 3 larger than GLM.     
Note, that since diffraction in GLM is Good-Walker type, our predicted elastic
and diffractive cross sections satisfy the Pumplin bound\cite{Pumplin},
$\sigma_{el}(s,b)+\sigma_{diff}(s,b) \leq \frac{1}{2}\sigma_{tot}$.
This bound does not aply to KMR, in which a significant part of its diffractive
cross section originate from Pomeron enhanced contributions.
\newline
4) An estimate of $S^2_H$ value
can be obtained, at an early stage of LHC operation, through a
measurement of the rate of central hard LRG di-jets production (a GJJG
configuration) coupled to a study of its expected rate in a non screened pQCD
calculation.
\newline
\vskip0.5cm
{\bf {\large Acknowledgments:}} 
This research was supported
in part by the Israel Science Foundation, founded by the Israeli Academy of Science
and Humanities, by BSF grant $\#$ 20004019 and by
a grant from Israel Ministry of Science, Culture and Sport and
the Foundation for Basic Research of the Russian Federation.



\end{document}